# Understanding and Interpretations of Quantum Mechanics

Dong Luo
South China University of Technology
luodong@scut.edu.cn

**Abstract**
Taking Heisenberg's and Schrödinger's theories of quantum mechanics as his case study, De Regt's contextual theory of understanding argues that recognizing qualitatively characteristic consequences of a theory T without performing exact calculations is a criterion for scientific understanding. From the perspective of this theory of understanding, the task of understanding quantum mechanics seems to have been achieved already or even finished. This appears to disagree with some physicists' attitude to the understanding of quantum mechanics in line with Richard Feynman's famous slogan "I think I can safely say that nobody really understands quantum mechanics." Moreover, if the task of understanding quantum mechanics has been finished already, there would be a conflict between the contextual theory of understanding of quantum mechanics and interpretations of quantum mechanics.

# INTRODUCTION

Philosophers posit various theories of scientific understanding. Some use case studies from the history of quantum mechanics to support their theories of scientific understanding (de Regt, 2017). Based on those case studies, they argue for certain criteria for the understanding of quantum mechanics and scientific understanding more generally. For them, there is a criterion for understanding of quantum mechanics, at least in certain contexts. Meanwhile, Richard Feynman's famous slogan "I think I can safely say that nobody really understands quantum mechanics" (Feynman, 1967, 129) won support among many physicists (e.g., Carroll, 2019). It appears that, for many physicists, they do not think they understand quantum mechanics, while for some philosophers, there are criteria for understanding quantum mechanics among physicists. This is weird. Why would philosophers think physicists understand quantum mechanics while many physicists say they do not?

Moreover, there are various interpretations of quantum mechanics since the birth of the theory, and still there is no agreement on which interpretation is the best for the understanding of quantum mechanics. If physicists have already reached agreements on the understanding of quantum mechanics, then what are efforts on interpretations of quantum mechanics for? If physicists, as Feynman's slogan indicates, have not achieved an understanding of quantum mechanics, then how could philosophers argue for a general theory of scientific understanding with case studies on the understanding of quantum mechanics from the history of quantum mechanics?

I will examine Henk de Regt's theory of scientific understanding and interpretations of quantum mechanics, and argue that the two conflict either on scientific understanding or on the understanding of quantum mechanics.

## DE REGT'S CONTEXTUAL THEORY OF SCIENTIFIC UNDERSTANDING

In his monograph *Understanding Scientific Understanding* (2017), Henk de Regt argues for a contextual theory of scientific understanding. According to the theory, "A phenomenon P is understood scientifically if and only if there is an explanation of P that is based on an intelligible theory T and conforms to the basic epistemic values of empirical adequacy and internal consistency " (de Regt, 2017, 92). The notion of intelligibility here is a rephrasing of the pragmatic understanding of a theory by de Regt. If scientists understand a theory, de Regt would say that the theory is intelligible to them (de Regt, 2017, 40). "A scientific theory T (in one or more of its representations) is intelligible for scientists (in context C) if they can recognize qualitatively characteristic consequences of T without performing exact calculations" (de Regt, 2017, 102). The criterion means that intelligibility or unintelligibility is not an intrinsic feature of theories, but is a context-dependent value.

To argue for the contextual theory of scientific understanding, de Regt distinguishes three levels of scientific activity: the macro-level which is practised by all scientists, the meso-level which is practised by scientists from different communities, and the micro-level which is practised by individual scientists. With this distinction de Regt thinks that, although aims on the macro-level are agreed on by all scientists, they could be articulated in different ways on the meso- or micro-level. For example, on the macro-level, all scientists will agree that they aim to produce knowledge that is supported by experience, but on the meso- or micro-level scientists from different communities and sometimes even scientists within a same community might have disagreements on how and how strongly scientific knowledge has to be supported by experience (de Regt, 2017, 90).

Following the distinction, de Regt argues that although achieving understanding is among the general (macro-level) aims of science, scientists in different historical periods or in different communities (on the meso- or micro-level) have quite different views about how precisely scientific understanding is to be achieved (de Regt, 2017, 90-91).

De Regt supported this context-dependent view of scientific understanding with case studies. One of his major case studies is the investigation of the debates among physicists in the 1920s on the intelligibility of two rival quantum theories, Heisenberg's matrix mechanics and Schrödinger's wave mechanics.

Heisenberg's theory was a highly abstract theory based on matrix theory that most physicists were unfamiliar with in the 1920s. It was intended to describe only relations between observable quantities, such as the frequencies and intensities of spectral lines emitted by atoms, and did not provide a visualizable model of the internal structure of atoms. In contrast to Heisenberg's theory, Schrödinger's wave mechanics was based on wave equations which were more familiar to physicists than matrix theory in the 1920s. It described the atom in terms of wave phenomena and suggested the possibility of visualizing atomic structure.

Supporters of the two theories debated which theory was superior. Schrödinger brought the notions of understanding and intelligibility to the debates and claimed that his wave mechanics was much better in providing a true understanding of quantum phenomena. De Regt thinks that Schrödinger argued for a position that only theories that are visualizable in space and time are intelligible and can give us the understanding of phenomena. And he believes that Schrödinger expressed a strong commitment to the view that visualization is a necessary condition for scientific understanding (de Regt, 2017, 4).

However, Heisenberg thought that what Schrödinger said about the intelligibility of wave mechanics scarcely makes any sense (de Regt, 2017, 243). Wolfgang Pauli argued further that although matrix mechanics appears less intelligible than wave mechanics, understanding it was just a question of becoming familiar with the new conceptual system of the matrix mechanics. Pauli believed that once physicists get familiar with the matrix theory, Heisenberg's matrix mechanics will also be intelligible (de Regt, 2017, 243-244). Heisenberg later adopted Pauli's views on intelligibility and claimed that "We believe to understand [*anschaulich zu verstehen*] a physical theory when we can think through qualitatively its experimental consequences in all simple cases and when we have checked that the application of the theory never contains inner contradictions" (Heisenberg, 1927,172; cited from de Regt, 2017, 244).

Although there were disagreements between supporters of the two theories, the competition ultimately led to their synthesis. Schrödinger's hope for a visualizable interpretation of quantum mechanics was not fulfilled for technical reasons and Heisenberg abandoned his radically abstract approach and reintroduced visualizable notions. The quantum mechanics that is accepted and taught today is a combination of matrix and wave mechanics（de Regt, 2017, section 7.3).

De Regt thinks that this debate in the history of quantum mechanics shows that standards of intelligibility and understanding may vary and change. He acknowledges that the debates inspire him to develop his contextual theory of scientific understanding （de Regt, 2017, 245).

Anyhow, from the perspective of the relationship between understanding and knowledge, it is not knowledge but intelligibility conceived as an ability that leads to understanding in de Regt's contextual theory of understanding. In other words, de Regt believes that the truth of a theory is not what leads to understanding.

## FEYNMAN'S SLOGAN AND DE REGT'S INTERPRETATION OF IT

Debates on the understanding of quantum mechanics played an essential role in the development of de Regt's theory. According to him, there was a criterion for the understanding of quantum mechanics, a synthesis of Schrödinger's and Heisenberg's criteria, after the matrix/wave-mechanics debates. Although the criterion is contextual, it indicates nevertheless that at least after the matrix/wave-mechanics debates physicists thought they understand quantum mechanics in some sense.

However, in 1967, Richard Feynman said "I think I can safely say that nobody really understands quantum mechanics" (Feynman, 1967, 129). This famous slogan was often quoted by physicists after Feynman and still wins support among many physicists (Carroll, 2019; Charap and Dombey, 2021; Baggott, 2020). Feynman's slogan was echoed by many later physicists and appears to contradict de Regt's finding that there was a criterion for the understanding of quantum mechanics at a certain historical period: If physicists agreed on the synthesized criterion for the understanding of quantum mechanics after the matrix/wave-mechanics debates, why would Feynman and many physicists claim that nobody understands quantum mechanics decades later? Did physicists change their criterion for the understanding of quantum mechanics in Feynman's time and quantum mechanics fails to satisfy the new criterion? Or are there developments in quantum mechanics that changed quantum mechanics such that the developed version of quantum mechanics no longer satisfies the criterion that emerged after the matrix/wave-mechanics debates?

Before considering these possibilities, it is necessary to show de Regt's response to Feynman's slogan:

> "... of course, he[Feynman] did not mean that nobody—not even experts in the field—understands the theory itself (like first-year physics students who do not understand quantum mechanics and hence fail their exam on the subject). Rather, he plausibly meant that even those who are familiar with the theory have trouble in seeing how one can understand the world if quantum mechanics is true." (de Regt, 2017, 45)

As we have stated at the beginning of the first section, according to de Regt to say that one understands a theory is to say the theory is intelligible to people. A theory is intelligible to certain groups of people at certain periods if those people know how to use the theory and more specifically know how to "recognize qualitatively characteristic consequences of T without performing exact calculations." Therefore, if de Regt believes that Feynman refers not to the understanding of quantum theory, then what "nobody" understands should concern quantum phenomena. Moreover, de Regt interprets the notion that nobody understands quantum phenomena as people having „trouble in seeing how one can understand the world if quantum mechanics is true."

De Regt did not explain why people would have trouble in understanding the world if quantum mechanics is true. However, he quotes another line from Feynman in the Introduction of his book:

> "Even the experts do not understand it the way they would like to, and it is perfectly reasonable that they should not, because all of direct, human experience and of human intuition applies to large objects." (Feynman *et al*. 1963–1965, 3:1 – 1)

A possible interpretation is that people have trouble understanding the world if quantum mechanics is true because people usually understand the world by direct human experience which applies to macro-objects, and the world understood by direct human experience contradicts what quantum mechanics says about the world. Anyway, Feynman's "nobody understands quantum mechanics" is interpreted by de Regt as nobody understands quantum phenomena.

## THE UNDERSTANDING OF QUANTUM PHENOMENA

If „nobody understands quantum mechanics" means that nobody understands quantum phenomena, then what does understanding a quantum phenomenon mean? De Regt's criterion for the understanding of a phenomenon is "A phenomenon P is understood scientifically if and only if there is an explanation of P that is based on an intelligible theory T and conforms to the basic epistemic values of empirical adequacy and internal consistency." Therefore, a quantum phenomenon P is scientifically understood if and only if there is an explanation of P that is based on an intelligible quantum mechanics theory T and conforms to the basic epistemic values of empirical adequacy and internal consistency.

From the discussion of the matrix/wave-mechanics debates we learn that Schrödinger believed, and Heisenberg and Pauli admitted, that wave mechanics is intelligible. Also, the debates led to an intelligible quantum mechanics theory which is a synthesis of matrix and wave mechanics. Therefore, the condition of "an intelligible quantum mechanics theory T" as the criterion for the understanding of a quantum phenomenon P is satisfied. The condition "conforms to the basic epistemic values of empirical adequacy and internal consistency" since the criterion would not pose a problem for both matrix and wave mechanics which are both empirically adequate and internally consistent. Therefore, to say a quantum phenomenon P is understood, one just needs to satisfy the condition that "there is an explanation of [the quantum phenomenon] P" which is based on an intelligible quantum theory.

What does it mean to be an explanation? There are debates about it. Here we do not need to examine various theories of scientific explanation in the philosophy of science. The way how de Regt understands or uses the notion of explanation in his contextual theory of scientific understanding is most relevant to our discussions here.

"An explanation is an attempt to answer the question of why a particular phenomenon occurs or a situation obtains, that is, an attempt to provide understanding of the phenomenon or the situation by presenting a systematic line of reasoning that connects it with other accepted items of knowledge (e.g., theories, background knowledge)" (de Regt, 2017, 24-25). Explanation also requires a pragmatic skill or ability to construct deductive arguments from the available knowledge to answer the question of why a particular phenomenon occurs or a situation obtains (de Regt, 2017, 24-25).

Therefore, to say that one has or gives an explanation of a quantum phenomenon P means that one has the skill or ability to construct deductive arguments from the available knowledge thato answer why the quantum phenomenon P occurs. "The available knowledge" for the construction of an understandable explanation includes an intelligible theory since the criterion for understanding is that "there is an explanation of P that is based on an intelligible theory T."

> "...the fact that the theory of matrix mechanics appeared unintelligible to many physicists hampered the construction of explanations to understand phenomena by means of this theory. Not only Schrödinger and most mainstream physicists, but even Bohr, Heisenberg, and Pauli had difficulties using matrix theory to explain and understand. By contrast, the more intelligible theory of wave mechanics yielded explanatory understanding of a wide variety of phenomena in a relatively straightforward manner. (Because of its initial unintelligibility—and the fact that it remains a counterintuitive theory that is difficult to master—many physicists adopted the positivist idea that quantum mechanics can furnish only description and prediction but no understanding of phenomena. This is a mistake, however.)" (de Regt, 2017, 91-92)

It is easy to see from the above quotation that de Regt thinks that, with intelligible wave mechanics physicists are already able to yield explanations of various quantum phenomena and obtain explanatory understandings of a wide variety of quantum phenomena. Moreover, de Regt claims that it is a mistake to think quantum mechanics cannot provide an understanding of quantum phenomena although quantum mechanics remains a counterintuitive theory.

What we obtain from the above discussion would lead to two possible consequences. First, if physicists are already able to yield explanations of quantum phenomena with an intelligible wave mechanics (as well as the quantum mechanics taught today that incorporates the intelligibility of Schrödinger's wave mechanics) and if they thus obtain understanding of quantum phenomena, then there is no further need to understand quantum phenomena. Also, de Regt's interpretation of Feynman's slogan of "nobody understands quantum mechanics" stated that people understand quantum mechanics theory. Now it would seem that people understand both quantum mechanics theory and quantum phenomenon. If both of them are understood, then what are the various interpretations of quantum mechanics (Copenhagen interpretation, many-worlds interpretation, etc.) about?

Second, if we agree with de Regt's interpretation that Feynman meant that nobody understands quantum phenomena when he said that nobody understands quantum mechanics, then the apparent contradiction between de Regt's study of the matrix/wave-mechanics debate and Feynman's slogan would be rephrased as: If physicists understand quantum phenomena based on an agreed synthesized criterion for the understanding of quantum phenomena after the matrix/wave-mechanics debates, why would Feynman and many physicists claim that nobody understands quantum phenomena decades later?

Is it because, with the developments of quantum mechanics in recent decades, people only now understand quantum mechanics? No! It is absurd to say that quantum mechanics is now understood whereas it was not 50 years ago. The conceptual framework of quantum physics remains as it was. Most applications of quantum mechanics (nuclear plants, medical scans, lasers, etc.) were understood 50 years ago.

Is it, finally, because a new or re-interpretation of the theory of understanding, such as de Regt's contextual theory of understanding, allows people to think that they understand quantum mechanics? If this were so, why are there still physicists who agree with Feynman that nobody understands quantum mechanics? Maybe it is because some physicists haven't got to know these new theories of understanding that were developed by philosophers. Anyhow, if we say that a new or re-interpretation of the theory of understanding makes quantum mechanics understandable, then, again, what are interpretations of quantum mechanics for?

## CONCLUSION

When people talk about understanding quantum mechanics, it is important to first know which aspects of quantum mechanics need to be understood. If we accept de Regt's contextual theory of scientific understanding, it seems that the task of understanding quantum mechanics has already been achieved or even finished as both quantum mechanics theory and quantum phenomena do not require to be further understood. If this is so, interpretations of quantum mechanics as one of the central issues in the philosophy of physics might be doomed to be meaningless.

There were complaints that quantum mechanics needs no interpretations (Fuchs & Peres, 2000; de Ronde, 2020). I'm not saying that I agree with this position, but I think there might be conflicts between de Regt's contextual theory of scientific understanding and the various efforts regarding interpretations of quantum mechanics.

**Acknowledgment** This research was conducted under the support of Chinese National Social Science Fund (No. 18CZX012).

## REFERENCES

Baggott, J. (2020). *Quantum Reality*. Oxford: Oxford University Press.

Carroll, S. (2019). Even Physicists Don't Understand Quantum Mechanics. *The New York Times*, Sep.7, 2019.
https://www.nytimes.com/2019/09/07/opinion/sunday/quantum-physics.html

Charap, J. and N. Dombey (2021). Quantum of solace: even physicists are still scratching their heads. *The Guardian*, Sep. 3, 2021. Quantum of solace: even physicists are still scratching their heads | Letters | The Guardian

de Ronde, C. (2020). Quantum Theory Needs No 'Interpretation' But 'Theoretical Formal-Conceptual Unity'. *arXiv*:2008.00321v1.

de Regt, H. W. (2017). *Understanding Scientific Understanding*. New York: Oxford University Press.

Feynman, R. (1967). *The Character of Physical Law*. Cambridge: MIT Press.

Fuchs, C.A. & Peres A. (2000). Quantum theory needs no "interpretation". *Physics Today* 53, 70-71.

Heisenberg, W. (1927). Über den anschaulichen Inhalt der quantentheoretischen Kinematik und Mechanik. *Zeitschrift für Physik* 43:172–198.